\documentclass[runningheads]{llncs}
\usepackage[roman]{parnotes}
\usepackage{graphicx}
\usepackage{hyperref}
\usepackage{multirow}
\begin{document}
\title{A Simple and Efficient Framework for Identifying Relation-gaps in Ontologies}
\titlerunning{Identifying Relation-gaps in Ontologies}
%
\author{Subhashree S \and
P Sreenivasa Kumar}
\authorrunning{Subhashree et al.}

%
\institute{Department of Computer Science and Engineering, Indian Institute of Technology - Madras, Chennai, India. 
\email{\{ssshree,psk\}@cse.iitm.ac.in}}
\maketitle              

\begin{abstract}
Though many ontologies have huge number of classes, one cannot find a good number of object properties connecting the classes in most of the cases. Adding object properties makes an ontology richer and more applicable for tasks such as Question Answering. In this context, the question of which two classes should be considered for discovering object properties becomes very important. We address the above question in this paper. We propose a simple machine learning framework which exhibits low time complexity and yet gives promising results with respect to both precision as well as number of class-pairs retrieved.\\
\end{abstract}
\section{Introduction}
In this work, we propose a novel and simple approach to identify relation-gaps in an ontology with the main focus of achieving low response-times. The goal is to find potential pairs of classes that could be connected by an object property but have not yet been connected i.e. they are relation-gaps in the ontology. Note that the focus is not on discovering the object properties which connect a \emph{given} pair of classes. While there are systems such as OntExt \cite{ontext} and DARO \cite{ngc} for the above task, our goal in this work is to identify the pairs of classes which could serve as input to these systems. Identifying relation-gaps becomes important because, feeding every non-connected class-pair as input to DARO would be inefficient. This is especially true in the case of large knowledge graphs (KGs) such as YAGO3 which has 488,469 classes but only 77 object properties \cite{yago3}. In order to add more object properties to YAGO3, one has to consider a huge number ($\approx 488469^2$) of class pairs, unless a better approach is devised.

\section{Related Works}
Prophet \cite{prophet} predicts pairs of classes to be connected by object properties, mainly in the NELL KG. Predicting links between nodes using the count of common neighbours between them is very popular in social network settings. Prophet bases its working upon this notion. Given a pair of classes, Prophet computes a score as the sum of common neighbours of all pairs of instances in the two classes, normalized by the number of instance pairs. The class pairs having a score above 10 are output by Prophet. The disadvantages of this approach are: (1)  If the given ontology is not rich enough, we cannot expect Prophet to output many new class-pair connections. (2) It has a high response-time as it considers every pair of instances in the given two classes and computes their common neighbours. In our experiments, we observed that Prophet (when we implemented it on a machine with 16 GB main memory) takes \emph{three hours} on an average to identify potential partners for one class in the DBpedia dataset. 
In our previous work \cite{ngc}, we had proposed a solution based on word embeddings for the problem of identifying relation-gaps. We claimed and experimentally proved that looking for common neighbours between two classes using \emph{external} sources leads to richer and more diverse connections in the KG. We used Word2Vec for this purpose as the word vectors learnt by the Word2Vec algorithm are such that two words which have high number of common neighbouring words have highly similar representations. This system has low response-time (around 5 seconds on a 32 GB main memory system). The major disadvantage of this system is that it does not give good results for very generic classes like ``Person". For ``Person", the system outputs classes such as ``Name", ``Year" and for more specific classes like ``Athlete", the system returns meaningful partners such as ``SportsLeague". (All the names in quotes are class names in DBpedia ontology.)

\section{Proposed Framework} 
We propose a machine learning framework for identifying relation-gaps in an ontology. The major goal of our system is to achieve low-response time. We design our features such that they do not rely upon the instances of the input classes as this tends to increase the runtime of the system. For example, we check for common neighbours between 2 given classes at the class-level while Prophet does this at an instance level. In our previous work \cite{ngc} we observed that the best results were given by three techniques - using Word2Vec, finding common neighbours and using the Adamic-Adar index. We also observed that the results given by our Word2Vec-based method were complementary to those given by the other 2 techniques. Hence in this work, we build an SVM classifier which takes these 3 quantities as its features. The features used are as below:
\emph{Common-Neighbours (CN):} This measure captures the number of shared neighbors between both the nodes. A neighbour of a class is a class that is already linked to it by an object property. Let $\Gamma(x)$ denote the set of neighbours of a node $x$. Then $cn_{xy}$= $|\Gamma(x) \cap \Gamma(y)|$.
\emph{Adamic-Adar Index (AA):} This index is similar to the above feature, but assigns more weight to the less connected neighbours \cite{adamicadar}. It is defined as $aa_{xy}$=$\sum_{z \epsilon \Gamma(x) \cap \Gamma(y) } \frac{1}{log|\Gamma(z)|}$.
\emph{GloVe embeddings:} In our previous work \cite{ngc} we had used Word2Vec vectors for generating relation-gaps. However, GloVe directly focuses on word co-occurrences over the available corpus and its embeddings relate to the probabilities that two words appear together. Since GloVe's mechanism is more directly associated with finding common neighbours based on their co-occurrences, we use GloVe embeddings\footnote{pre-trained embeddings of 100 d - \scriptsize{ \url{http://nlp.stanford.edu/data/wordvecs/glove.6B.zip}}} in this work. 

\section{Experiments and Results}\label{exp}

We consider DBpedia version 2016-10 for extracting the positive instances of our training data. There are 1105 object properties, and 708 among them have domain and range assigned\footnote{obtained by querying the DBpedia SPARQL endpoint  on 31st May 2020}. Among these, we eliminate duplicate domain-range connections and obtain 335 domain-range pairs as positive instances. In order to obtain negative instances, we manually identify 279 class pairs in the DBpedia ontology as those which cannot be related by any object property (for e.g. Cheese and Mountain). We test our classifier on six ontologies (details are in Table \ref{testonto}) - four ontologies have been built by our own research group\footnote{DSA, WM, MHBT, HP ontologies - \scriptsize{\url{https://sites.google.com/site/ontoworks/ontologies}}} and two are from public repositories\footnote{Pet ontology - \scriptsize{\url{http://www.cs.man.ac.uk/~horrocks/ISWC2003/Tutorial/}}\\ PP ontology - \scriptsize{\url{https://sites.google.com/site/ppontology/}}}. We have chosen the test ontologies such that: a major fraction of the object properties do not have their domain and range specified (HP, Pet, WM ontologies); large number of individuals are present (PP, MHBT, DSA ontologies). These characteristics have a direct impact on the working of our competing systems. We manually evaluate the positive class-pairs newly-found by our proposed approach, for each ontology. Three ontology engineers (non-authors) checked whether the pairing of classes makes sense. They were asked to mark the pair as: \emph{correct} or \emph{incorrect}. For a class-pair to be counted as correct, two out of the three evaluators should have agreed on it. Table \ref{results} shows sample class-pairs generated and time taken by the proposed system (for the entire ontology, when run on a system with 16 GB main memory) and the precision value (ratio of correct class-pairs to the total class-pairs) for all the three systems - the proposed approach, our earlier work called here as WV-based\cite{ngc} and Prophet\cite{prophet}. List of all class-pairs generated can be seen in the project web page\footnote{\scriptsize{\url{https://sites.google.com/site/ontoworks/projects}}}. From Table \ref{results}, it can be seen that the proposed system significantly outperforms the other two systems with respect to the number of relation-gaps identified. Prophet generates results only for ontologies which have high number of instances (MHBT and PP) as its mechanism is based on finding common neighbours at the instance level. Though the DSA ontology has high number of individuals, Prophet fails to produce results because it lacks many relation instances. For ontologies which don't have domain and range specified for many of the object properties (HP, Pet and WM), the features based on common-neighbours and Adamic-Adar index fail to predict any result. However, the GloVe-based feature of our model plays a major role in such input cases to give good results. Though the WV-based system produces results for all ontologies, it generates lesser number of results compared to the currently proposed system for most of the cases. 
\begin{table*}
  \caption{Specifications of Test Ontologies}
  \label{testonto}
  \begin{tabular}{p{4.5cm}|c|c|c|c}
     \hline\noalign{\smallskip}
     Dataset & Classes & Individuals & Object Properties (OP) & OP w/o domain and range\\
     \noalign{\smallskip}\hline\noalign{\smallskip}
Data Struct. and Algo. (DSA) & 107 & 154 & 26 & 2\\
WikiMovie (WM) & 35 & 104 & 14 & 5\\
Mahabharata (MHBT) & 22 & 249 & 33 & 11\\
Harry Potter (HP) & 17 & 12 & 5 & 5\\
People \& Pets (Pet)&  60 & 21 & 14 & 13\\
Plant Protection (PP) & 92 & 548 & 15 & 0\\
     \noalign{\smallskip}\hline
  \end{tabular}
\end{table*}
\begin{table*}
  \caption{Sample pairs, time (proposed system) and precision (correct/produced pairs)}
  \label{results}
  \begin{tabular}{l|l|c|c|c|c}
     \hline\noalign{\smallskip}
   \multirow{2}{*}{Dataset}  &   \multirow{2}{*}{Sample results by the proposed approach}  & Time-taken & \multicolumn{3}{c}{Comparison of Precision}\\
     
      &   & (in seconds) & Proposed & WV-based & Prophet \\
     \noalign{\smallskip}\hline\noalign{\smallskip}
DSA & (Graph\_Traversal, Undirected\_Graph) & 9 & 165/176(0.94) & 125/136(0.92) & no results\\
WM & (Film\_producer, Genres); (Actor, Language) & 6 & 127/127(1) & 246/246(1) & no results\\
MHBT & (Pandava, Kaurava); (Events, Places) & 6 & 39/41(0.95) & 5/5(1) & 16/16(1)\\
HP & (Gryffindor, Slytherin) & 5 & 21/22(0.95) & 8/10(0.8) & no results\\
Pet & (pet+owner, pet); (truck, bicycle) & 6 & 165/176(0.94) & 123/130(0.95) & no results\\
PP & (Disorder, Abnormality); (Pest, Pesticide) & 8 & 175/178(0.98) & 141/172(0.82) & 14/14(1)\\
     \noalign{\smallskip}\hline
  \end{tabular}
\end{table*}

\section{Conclusions}
 In this paper, we have proposed a low response-time framework for identifying relation-gaps in an ontology. Using the insights gained from our previous work, we have carefully picked the most useful features to build our classifier. The proposed system gives low response-time on all the tested ontologies mainly because the chosen features are not dependent on the number of class instances. The proposed system substantially beats the competing systems with respect to number of class-pairs returned while maintaining very good precision. 

%
%

%
%
%
 \bibliographystyle{splncs04}
 \bibliography{main}

\end{document}